\documentclass[a4paper]{book}
\usepackage{nano2cmr}
\usepackage[numbers,sort&compress,sectionbib]{natbib}

\usepackage{hyperref}
 \usepackage{amssymb}
 \usepackage{graphicx}

\begin{document}

\pnum{}
\def\tit{Quantum dynamics of a domain wall in the presence of
  dephasing} \ttitle{\tit}
\tauthor{Claudio Castelnovo, Mark I. Dykman, Vadim N. Smelyanskiy,
  Roderich Moessner, {\em Leonid P. Pryadko}}

\ptitle{\tit}

\pauthor{Claudio Castelnovo$^{1}$, Mark I. Dykman$^{2}$, Vadim
  N. Smelyanskiy$^{3}$, Roderich Moessner$^{4}$, {\em Leonid
    P. Pryadko}$^{5}$}

\affil{$^{1}$~T.C.M. Group, Cavendish Laboratory, University of
  Cambridge, J. J. Thomson Avenue, Cambridge CB3 0HE, U.K.
\\ $^{2}$~Department of Physics and
  Astronomy, Michigan State University, East Lansing, Michigan 48824,
  USA
\\ $^{3}$~Google Inc., Venice,
  California 90291, USA
\\ $^{4}$~Max-Planck-Institut f\"{u}r Physik komplexer Systeme,
  01187 Dresden, Germany
\\ $^{5}$~Department of Physics \&
  Astronomy, University of California, Riverside, California 92521,
  USA}

\begin{abstract}
  {We compare quantum dynamics in the presence of Markovian
    dephasing for a particle hopping on a chain and for an Ising
    domain wall whose motion leaves behind a string of flipped spins.
    Exact solutions show that on an infinite chain, the transport
    responses of the models are nearly identical.  However, on
    finite-length chains, the broadening of discrete spectral lines is
    much more noticeable in the case of a domain wall.}%
\end{abstract}

\begindc 

\index{Castelnovo, C.}
\index{Dykman, M. I.} 
\index{Smelyanskiy, V. N.}   
\index{Moessner, R.}
\index{Pryadko, L. P.}

\section*{Introduction}

The role of dephasing on the time evolution of a quantum mechanical
system is a fundamental issue in the study of open quantum systems.

We pose the question what happens if the object moving around is not a
simple pointlike particle, but rather an emergent quasiparticle which
acts as a source of an observable emergent gauge field.  One example
is the monopole and Dirac string
dynamics\cite{Jaubert-Holdsworth-2009,Wan-Tchernyshyov-2012} in spin
ice\cite{Castelnovo-Moessner-Sondhi-2012}.  As the full motion in a
disordered spin background is beyond the scope of a first pass at this
problem, we consider a simplified setting where the motion takes place
in one dimension; we contrast the cases of a free particle and one
with a string attached, in the form of a domain wall in an Ising
system.  Experimentally, this question corresponds, e.g., to the
observation of a Villain mode in a quasi--one-dimensional
magnet\cite{Villain-1975,Nagler-Buyers-Armstrong-Briat-1982}.

Our central results are the following.  The first is a technical one,
namely that we can solve both cases (particle and domain wall motion)
for a one-dimensional model subject to a locally uncorrelated
Markovian dephasing bath.  Secondly, this solution demonstrates that,
for unstructured motion in one dimension, the two cases differ only
weakly, in the sense that the difference between the two is
considerably smaller than the difference between either and the fully
coherent time evolution.  In particular, linear transport responses in
the presence of a linear density gradient or a uniform field are
identical for the two cases.  Thirdly, we notice that this is no
longer the case when considering finite-length chains.  Here, the
discrete energy spectrum is broadened considerably more strongly for
the case of domain walls; this can be qualitatively understood as the
enhanced fragility of the interference of a domain wall with itself as
it does a round trip on the finite lattice to establish the standing
wave.

\section{Models}

The two models we solve both describe single-body one-dimensional
hopping in the presence of Markovian dephasing uncorrelated between
the sites, written in terms of the density matrix with components
$\rho_{ab}$,
\begin{equation}
  \label{eq:model-general}
\dot \rho_{ab}=-i[H,\rho]_{ab}-\Gamma_{a-b} \,\rho_{ab}, \quad
(\text{no summation!})
\end{equation}
where dephasing rates for all off-diagonal elements of the density
matrix are equal in the case of a particle (P), 
while they grow linearly with the distance from the diagonal in the
case of a domain wall (DW), 
\begin{equation}
  \label{eq:gamma-all}
  \Gamma_{s}^{({\rm P})}=\gamma\,(1-\delta_{s,0}),\qquad 
  \Gamma_{s}^{({\rm DW})}=\gamma\,|s|.
\end{equation}
In Eq.~(\ref{eq:model-general}),  $H=H_0$ with the matrix elements
\begin{equation}
\label{eq:ham-hopping}
(H_0)_{ab}=-{\Delta\over2} (\delta_{a,b+1}+\delta_{a+1,b})
\end{equation}
is the usual hopping Hamiltonian, $\delta_{a,b}$ is the Kronecker
symbol, and the parameters $\Delta$ and $\gamma$ respectively denote
the half band width and the dephasing rate.  

Formally, Eqs.~(\ref{eq:model-general}) with 
$\Gamma_s^{(\rm P)}$ of Eq.~(\ref{eq:gamma-all}) can
be considered a Lindblad equation\cite{lindblad-76} for particle
hopping, where each site has its own bath coupled to its
occupation number.  It describes universal long-time dephasing physics
valid in the limit where both the bath cutoff frequency $\omega_c$
(maximum frequency of a bath mode) and the bath temperature
$\beta^{-1}$ are large compared to the hopping band width $2\Delta$.

Similarly, with $\Gamma_s^{(\rm DW)}$ of Eq.~(\ref{eq:gamma-all}),
these equations describe dynamics in a single-DW sector of an Ising
spin chain in the presence of the transverse field $\Delta$, and
independent fluctuating longitudinal magnetic fields.  Moving the
domain wall by $|a-b|$ positions requires flipping $|a-b|$ spins,
which increases the dephasing rate for the matrix element $\rho_{ab}$.

\section{Results}

On an infinite chain, we use the translational symmetry and define the
Fourier transformation for the center of mass,
\begin{equation}
  \rho_{ab}=\int {dK\over 2\pi} e^{i KR}e^{i\pi s/2}\phi_s(K,t),\quad
  s\equiv a-b,
  \label{eq:slow-K}
\end{equation}
where $R\equiv (a+b)/2$ and the phase factor $e^{i\pi s/2}$ makes
explicit the reflection symmetry, $s\to-s$.  The densities
$\phi_s(K,t)$ satisfy a 1D Schr\"odinger equation with
hopping $u_K=\Delta\sin (K/2)$ and an imaginary on-site potential
$-i \Gamma_s$.

(\textbf{i}) With $K=0$, densities $\phi_s(0,t)$ at different $s$ are
independent from each other and, for $s\neq0$, decay to zero with
rates $\Gamma_s$.  It is also easy to see that stationary solutions of
Eq.~(\ref{eq:model-general}) with the diagonal density
$n_s\equiv \rho_{ss}$ in the form of a polynomial of degree $\ell\ge 0$
have all off-diagonal matrix elements with $|a-b|>\ell$ zero.  In
particular, with $n_s$ linear, the stationary solution $\rho_{ab}$ is
tri-diagonal, which makes the linear diffusive transport of particles
and DWs identical.

\begin{figure}[htbp]
  \centering
  \includegraphics[width=0.9\columnwidth]{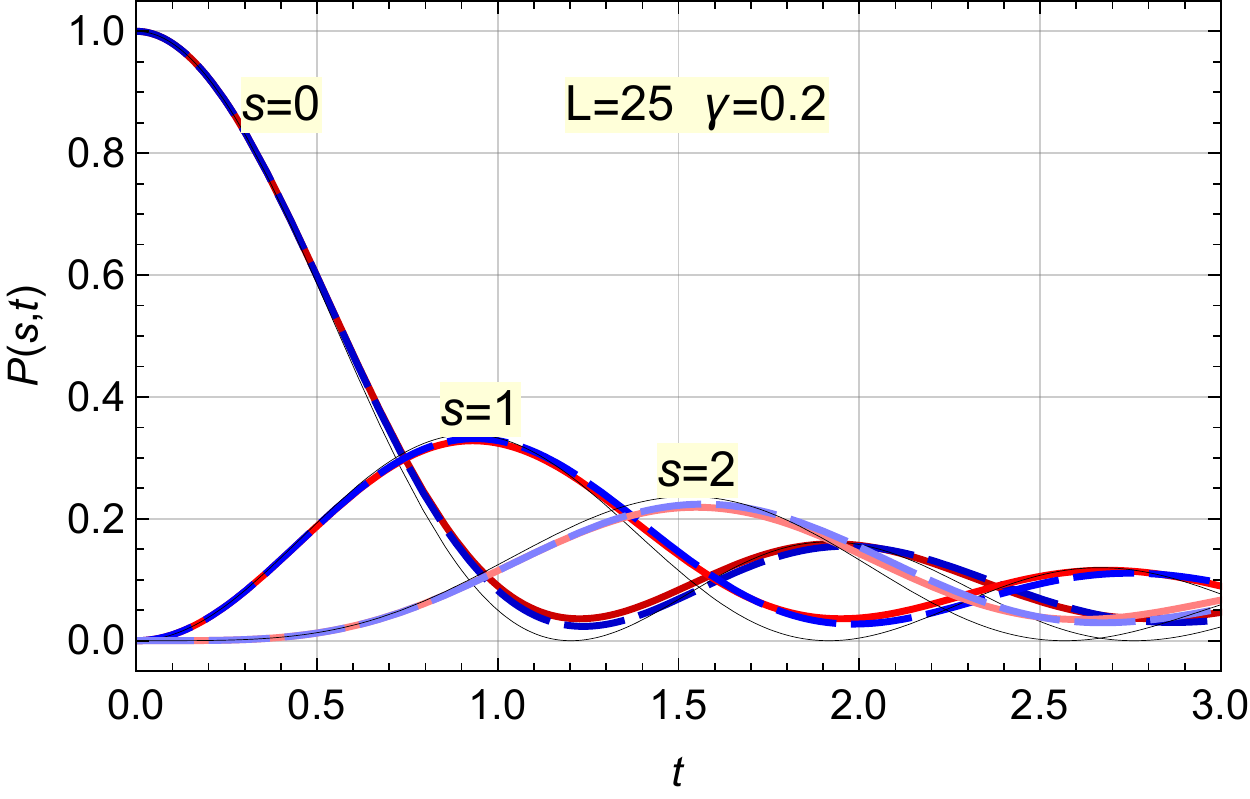}\vskip-1em
  \caption{Time dependence of the probabilities $P(s,t)$, $s=0,1,2$,
    for a particle (red solid lines) and for a DW (blue dashed lines)
    with $\Delta=2$ and dephasing $\gamma=0.2$.  Thin black lines show
    the corresponding results at $\gamma=0$,
    $P(s,t)=J_s^2(\Delta\, t)$, where $J_s(z)$ is the Bessel function
    of order $s$.  Even for such a relatively large $\gamma$, there is
    little difference between a particle and a DW.}
  \label{fig:ptcmp}
\end{figure}

(\textbf{ii}) We obtained explicit solutions for Laplace-transformed densities
$\psi_s\equiv {\psi}_s(K,p)$, with the initial condition
$\phi_s(K,t=0)=\delta_{s,0}$ of a classical purely-diagonal
density matrix:
\begin{eqnarray*}
\lefteqn{\quad  \psi^{({\rm
  P})}_{s}={e^{-i|s|\pi/2}\bigl[y\bigl({p+\gamma\over 2
  u_K}\bigr)\bigr]^{|s|}\over
  [(p+\gamma)^2+4 u_K^2]^{1/2}-\gamma},\;\, 
  y(t)\equiv (1+t^2)^{1/2}-t,} \qquad
  \qquad \qquad\qquad  & & \hskip2.5in\quad\quad  \\
\lefteqn{\quad\psi_s^{({\rm DW})}=
 e^{- i|s|\pi/2}{I_{p/\gamma+|s|}(z)\over \gamma z \,
  I_{p/\gamma}'(z)}, \quad z\equiv 2u_K/\gamma,}\qquad \qquad \qquad
  \qquad & & \hskip1.5in
\end{eqnarray*}
where $I_\nu(z)$ is the modified Bessel function of the first kind,
and $I_\nu'(z)=I_{\nu+1}(z)+(\nu/z) I_\nu(z)$ is its derivative.  At
$s=0$, these quantities are related to the dynamic structure factor
$S(K,\omega)=\psi_0(K,\epsilon+i\omega)$, given by the spatial and
temporal Fourier transform of the probability $P(s,t)$ to travel $s$
sites in time $t$.  While functional forms differ for the two cases,
plots of $S(K,\omega)$ at $\gamma/\Delta< 1/10$ are remarkably
similar (not shown due to space constraint).  Also similar are the
probabilities $P(s,t)$, see Fig.~\ref{fig:ptcmp}.

\begin{figure}[htbp]
  \centering
  \includegraphics[width=0.85\columnwidth]{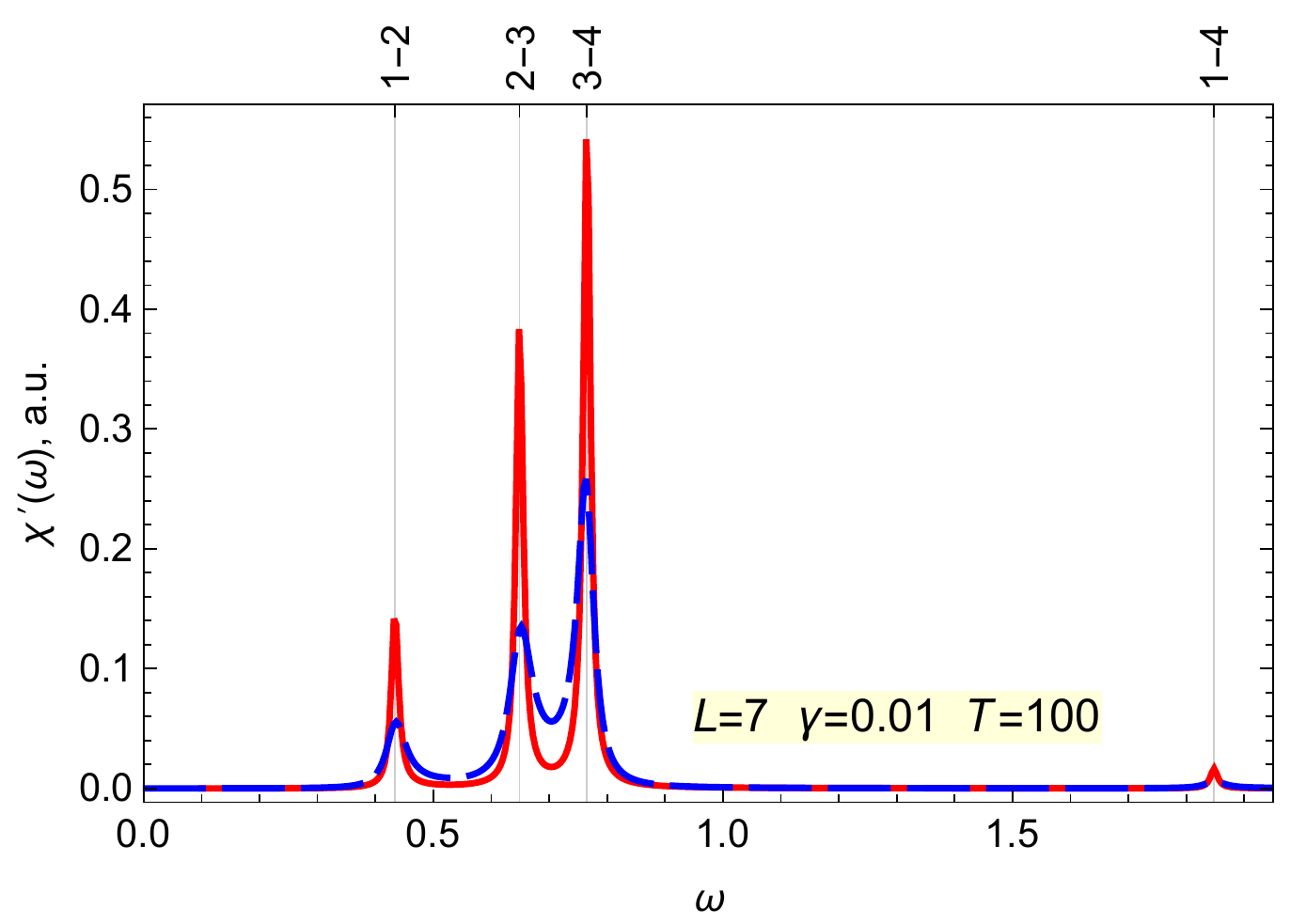}\\
  \includegraphics[width=0.85\columnwidth]{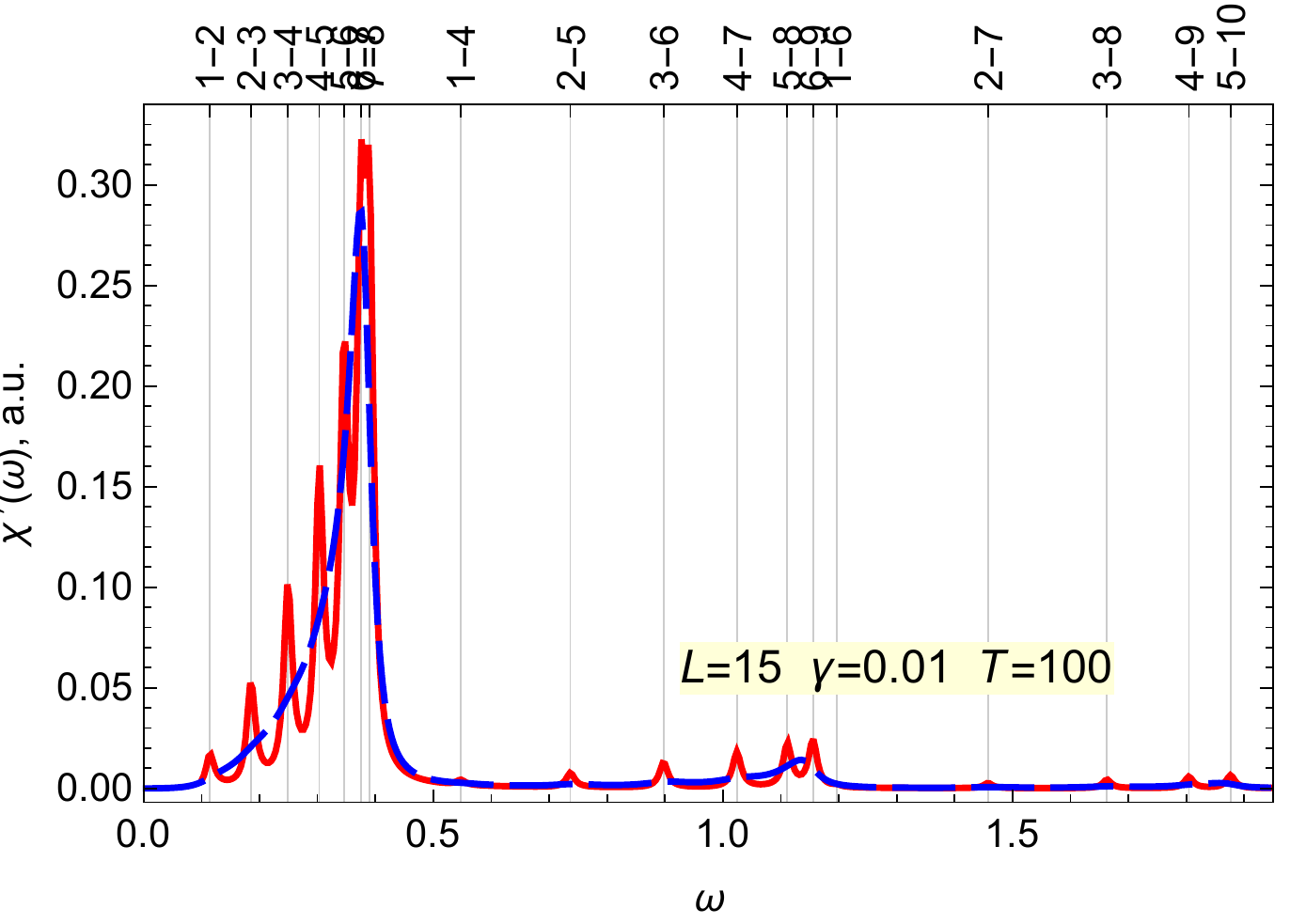}\vskip-1em
  \caption{Real part of the frequency-dependent susceptibility,
    $\chi'(\omega)$, for a particle (red solid lines) and a DW (blue
    dashed lines) on chains of length $L=7$ and $L=15$ as indicated,
    with half band width $\Delta=2$ and dephasing $\gamma=0.01$.
    Vertical grid lines mark allowed transitions between the discrete
    energy levels in the absence of dephasing (only one level pair is
    shown for each line).  While discrete lines for a particle are
    well resolved in both cases, they are suppressed entirely for a DW
    on the longer chain.  }
  \label{fig:pwcmp}
\end{figure}

(\textbf{iii}) In spite of these similarities, finite-frequency
responses on finite chains look different in the two cases.  Indeed, a
bound state is formed when a wave function interferes with itself;
such an interference requires off-diagonal matrix elements of
$\rho$.  Solving linearized versions of
Eqs.~(\ref{eq:model-general}) with a harmonically
modulated linear potential, and assuming the unperturbed thermal
density matrix $e^{-\beta H_0}$, we analyzed the average conductance
$\chi(\omega)$ [Fig.~\ref{fig:pwcmp}].  At small $\gamma$, the
corresponding real part $\chi'(\omega)$ has a series of resonant peaks
at $\omega_{mn}=E_m-E_n$, where symmetry requires $m-n$ to be odd, and
$E_m=-\Delta\cos k_m$, $k_m=\pi m/(L+1)$, are the energy levels of the
chain (\ref{eq:ham-hopping}) of length $L$, $m=1,2,\ldots, L$.  In the
case of a particle on such a chain, we find the width of each peak to
be $\gamma (L-1)/(L+1)$, while for a DW on a chain with $L$ allowed
positions, peak widths scale as $\gamma\,{\cal O}(L)$, as would be
expected on general grounds.


\section{Conclusions}
Quantum mechanics teaches us that particles can behave as waves, and
waves as particles.  For weakly-interacting particles, this is
described by second quantization.  A superficially similar
correspondence also exists outside of the perturbative sector where a
topological defect can be often viewed as a particle, its motion
described by the Schr\"odinger equation.  Some of the examples are
dislocations in lattices, vortices in 2D superfluids or
superconductors, and various soliton-like defects in 1D systems.
Experimentally observed quantum manifestations of such objects include
position uncertainty and quantum delocalization of lattice defects,
quantum tunneling of vortices and magnetic domain walls, and quantum
transport in conducting polymers.

Our main conclusion is that this analogy between collective
excitations and particles is not universal.  Environment can severely
limit the quantum behavior of such excitations.

\acks This work was supported in part by the ARO grant
W911NF-14-1-0272, the NSF grant PHY-1416578, and EPSRC grants
EP/K028960/1 and EP/M007065/1. 

\bibliographystyle{apsrevnourl}

\end{document}